\documentclass[
reprint,
superscriptaddress,
amsmath,amssymb,aps, pra,
floatfix, 10 pt]{revtex4-1}
\usepackage{multirow}
\usepackage{makecell}
\usepackage{graphicx}
\usepackage{dcolumn}
\usepackage{bm}
\usepackage[mathlines]{lineno}
\usepackage{graphicx}
\usepackage{bm}
\usepackage{multirow}
\usepackage{blindtext}
\usepackage{appendix}
\usepackage{graphicx}
\usepackage{amsmath}
\usepackage{xcolor}

\usepackage{longtable}
\usepackage[utf8]{inputenc}

\begin{document}
\preprint{APS/123-APS}

\title{Towards full understanding the physics of heavy ion induced inner shell ionization}
\author{Manpreet Kaur}
\affiliation{Department of Physics, University Institute of Sciences, Chandigarh University, Gharuan, Mohali, Punjab 140413, India}
\author{Soumya Chatterjee}
\affiliation{Department of Physics, Brainware University, Barasat, Kolkata - 700125, India.}
\author{T. Nandi}
\email[email:~]{nanditapan@gmail.com}
\affiliation{Department of Physics, Ramakrishna Mission Vivekananda Educational and Research Institute, PO Belur Math, Dist Howrah 711202, India}

\date{\today}

\begin{abstract}
\noindent
The complex physics of inner shell ionization of target atoms by heavy ion impact has remained only partially solved for decades. Recently, agreement between theory and experiment has been achieved by considering inner shell ionization of target atoms due to projectile electron capture in addition to direct Coulomb ionization including multiple ionization effects. A thorough investigation exhibits such a picture only if the atomic parameters of the target atoms are correct. In fact, the theoretical approach is found to be right, but the problem arises with the faulty atomic parameters. Furthermore, we show that fluorescence yields play a major role among the atomic parameters. We explore such a powerful method that enables us to measure the correct and accurate fluorescence yields for almost every element in the periodic table. As per our present knowledge, this in turn not only solves the said complex issue fully but also makes the PIXE analysis more reliable and accurate using both light and heavy ions.
\end{abstract}

\maketitle

\section{Introduction}
\indent The inner-shell ionization of heavy atoms by energetic positive ions \cite{brandt1973dynamic} is an important field of research for many applications including the calibration and efficiency measurement of x-ray detectors \cite{tribedi1992efficiency}, particle/proton-induced x-ray emission (PIXE) analysis \cite{JOHANSSON1976473, satoh2015development,lapicki2014werner}, medical \cite{lewis1997medical}, industrial \cite{JOHANSSON1976473, INDUSTRY}, atomic and nuclear physics \cite{dyson2005x}, particle physics \cite{undagoitia2015dark}, astrophysics \cite{deprince2020k}, plasma diagnostics \cite{sharma2016experimental} and testing the theoretical descriptions \cite{brandt1981energy} since decades \cite{khan1965studies}. However, disagreements are often found between the experiments and the currently available theories even in simple K-shell ionization by light-ion impact \cite{lapicki1989cross}. This problem with heavy-ion-induced K-shell ionization \cite{de2005k} is more severe, as this problem is theoretically more complicated than that with light ions due to multiple ionizations created in the outer shells of target atoms \cite{msimanga2016k}. To solve such a complex issue, we take up the case of the K-shell ionization study of copper, zinc, and germanium by the impact of silicon ions with energies of 3-5 MeV/u \cite{chatterjee2021significance}. We notice that direct Coulomb ionization (DCI) along with multi-ionization (MI) effects are much lower than those of the experimental scenarios. If we consider the knowledge that the mean charge state of the projectile ions inside the solid target is considerably higher than that outside the target \cite{sharma2016x} and then make use of the charge state distributions of the heavy ions inside the target in calculating the K-shell ionization by electron capture by the heavy projectile ions in addition to DCI with MI, we see good agreement \cite{chatterjee2021significance}. This idea has been tested through the L-shell ionization of osmium atoms by fluorine ions. Here, we see a good agreement between theory and experiment quite well, but a better agreement is achieved by tuning the atomic parameters (AP) \cite{chatterjee2022understanding}. Next, we move ahead to examine the same through M-shell ionization of tantalum, tungsten, lead,\href{\href{}{}}{} and bismuth by impact of neon ions \cite{kaur2023understanding}. Although all of these experiments were performed in the same laboratory \cite{GORLACHEV201919}, the said theoretical approach agrees well only with lead and bismuth. 
\par
We learn from our ongoing studies \cite{chatterjee2021significance, chatterjee2022understanding, kaur2023understanding} that heavy ion-induced ionization of target atoms is complex due to the major consequence of the electron capture process and the minor effect of the multi-ionization phenomenon. However, the evidence noticed in the L- and M-shell studies \cite{chatterjee2022understanding, kaur2023understanding} indicate that accounting for the electron capture process and the multi-ionization phenomenon is not always sufficient to resolve the long-lasting discrepancies prevalent between the experiments and theories for heavy ion-induced ionization processes. Furthermore, we mention here that the inner shell ionization cross-section by heavy-ion impact is much higher than that by light-ion impact. Thus, heavy-ion-induced PIXE (HIPIXE) is more efficient than light-ion-induced PIXE (LIPIXE). However, HIPIXE remains far from being utilized due to the issue mentioned above (for example, \cite{siegele2019heavy}). \textcolor{blue}{Hence, resolving such issue is highly indispensable.}  \\
\section{Methods of determining the fluorescence yields}
\begin{figure}
\centering
\includegraphics[width=8.7cm,height=9.0cm]{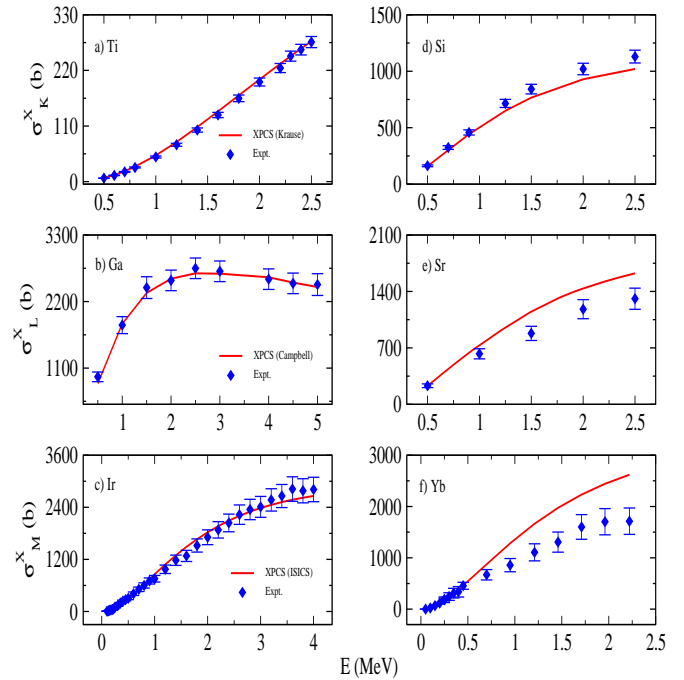}
\vspace{-20pt}
\caption{Mixed scenario of agreement and disagreement is observed while experimental x-ray production cross-sections are compared with the corresponding theoretical values from ECPSSR by taking atomic parameters (AP) from \citet{campbell2003fluorescence, BATIC20132232}. The comparison is displayed for K x-ray production cross-sections of Ti and Si in (a) and (d), L x-ray production cross-section of Ga and Sr in (b) and (e), and M x-ray production cross-section of Ir and Yb in (c) and (f), respectively. The cause of this mismatch is nothing but the correct AP for Ti, Ga, and Ir on one side and the incorrect AP for Si, Sr, and Yb on the other side.}
\label{Match-Mismatch}
\vspace{-10pt}
\end{figure}

\indent To explore the real cause of the above-mentioned issue, we conducted a thorough survey on proton-induced inner shell ionization processes, where the influence of the electron capture process as well as the multiionization phenomenon is almost nil. We have made a thorough comparison between experimental and theoretical x-ray production cross-sections (XPCS) due to the creation of a vacancy of K- \cite{tribedi1992k, SIMON-PhysRevA.39.3884, paul1989fitted}, L- \cite{ orlic1994experimental,miranda2014experimental}, and M-shell \cite{BALWINDER} in many elemental targets due to proton impact. Theoretical calculations are performed with \textcolor{red}{one} successful theoretical approach known as ECPSSR to include the effect of energy loss, Coulomb deflection, perturbed stationary state, and relativistic correction during ion impacts \cite{brandt1981energy} using the ISICSoo code \cite{BATIC2012398}. Figure \ref{Match-Mismatch} gives a representative picture of the experimental K, L, and M X-ray production cross-sections of some target atoms by the impact of a proton compared to the predictions from ECPSSR. This figure displays striking evidence that even the proton-induced ionization phenomenon is not well represented by the theory for all elements, regardless of whether the x-ray production study is concerned with K- or L- or M-shell ionization of the target atoms. To identify the origin of such mixed scenarios, let us first see how we determine the ionization cross-sections (ICS) from the experimentally obtained XPCS so that the theoretical estimates can be compared with the corresponding experimental results.\\
\indent Theoretically K shell ionization cross-section (ICS) $\sigma^I_{K}$ is related to K x-ray production cross-section ($\sigma_K^x$) as follows
\begin{equation}
\sigma^I_{K}= \frac{\sigma^x_{K}}{\omega_K},\label{eqn:Sig-Kxx}
\end{equation}
where $\omega_K$ is the K x-ray fluorescence yield. Similarly, the $L_i$ ($i=1,2,3$) subshell ionization cross sections $\sigma_{L_i}$ are related to the concerned $L$ x-ray production cross sections. 
To start with, $\sigma_{L_1}$ is given by

\begin{eqnarray}
\sigma_{L_1}= \frac{\sigma_{L_{\gamma_{2+3}}}}{{\omega_1}{F_{\gamma_{2+3,1}}}},\label{eq:L1}
\end{eqnarray}
\noindent where $\sigma_{L_{\gamma_{2+3}}}$ is the XPCS for $\gamma_{2+3}$ lines, ${F_{\gamma_{2+3,1}}}$ is the fraction of the radiative transition rate for ${L_{\gamma_{2+3}}}$ and the sum of all the transition rates due to x-ray emissions terminating in the $L_1$ subshell and $\omega_1$ is the fluorescence yield of all the x-ray emissions terminating in the $L_1$ subshell. Similarly, the $\sigma_{L_2}$ is obtained by
\begin{eqnarray}
\sigma_{L_2}= \frac{\sigma_{L_{\gamma_{1+5}}}}{{\omega_2}{F_{\gamma_{1+5,2}}}}-\sigma_{L_1}f_{12},\label{eq:L2}
\end{eqnarray}
\noindent where $\sigma_{L_{\gamma_{1+5}}}$ is the XPCS for $\gamma_{1+5}$ lines, ${F_{\gamma_{1+5,2}}}$ is the fraction of the radiative transition rate for ${L_{\gamma_{1+5}}}$ and the sum of all the transition rates due to x-ray emissions terminating in the $L_2$ subshell, $\omega_2$ is the fluorescence yield of all the x-ray emissions terminating in the $L_2$ subshell and $f_{12}$ is the Coster-Kronig (CK) yield for the transition $L_2$ to $L_1$. Lastly, the  $\sigma_{L_3}$ is derived by
\begin{eqnarray}
\sigma_{L_3}= \frac{\sigma_{L_{\alpha_{1+2}}}}{{\omega_3}{F_{\alpha_{1+2,3}}}}-\sigma_{L_1}(f_{12}f_{23}+f_{13})-\sigma_{L_2}f_{23},\label{eq:L3}
\end{eqnarray}
\noindent where $\sigma_{L_{\alpha_{1+2}}}$ is the XPCS for $\alpha_{1+2}$ lines, ${F_{\alpha_{1+2,3}}}$ is the fraction of the radiative transition rate for ${L_{\alpha_{1+2}}}$ and the sum of all the transition rates due to x-ray emissions terminating in the $L_3$ subshell and $\omega_3$ is the fluorescence yield of all the x-ray emissions terminating in the $L_3$ subshell.  $f_{13}$ and $f_{23}$  are CK yields for the $L_3$ to $L_1$ and $L_3$ to $L_2$ transitions, respectively. \\
\indent
\section{Results and discussions}
We have compared proton-induced XPCS for K, L, and M x-ray emissions between ECPSSR calculations and experiments where data are available. Some representative cases are shown in Fig.\ref{Match-Mismatch}. Here, we observe that good agreement is achieved for titanium K x-rays, but a departure is seen for that of silicon. Similarly, for the L x-ray production cross-sections, a good consonance is found for gallium, but that is not so for strontium. In the case of M x-rays, good compatibility is seen for iridium, but certain departure is noticed for ytterbium. Note that AP are energy independent, while the ICS are energy dependent. Thus, in the cases where the agreement is good with the energy variation, both ICS and AP are correct entities \textcolor{red}{within their respective uncertainties.} However, in cases of disagreements \textcolor{red}{three cases may arise: (i) both ICS and AP are incorrect, (ii) only the ICS is incorrect and (iii) only the AP is incorrect.} However, many works have doubted the AP in various contexts (for example, \citet{jones2007x}, \citet{chatterjee2022understanding}) and thus we first consider the case of correct ICS and faulty AP.

\begin{figure}
\centering
\includegraphics[width=8.7cm,height=9cm]{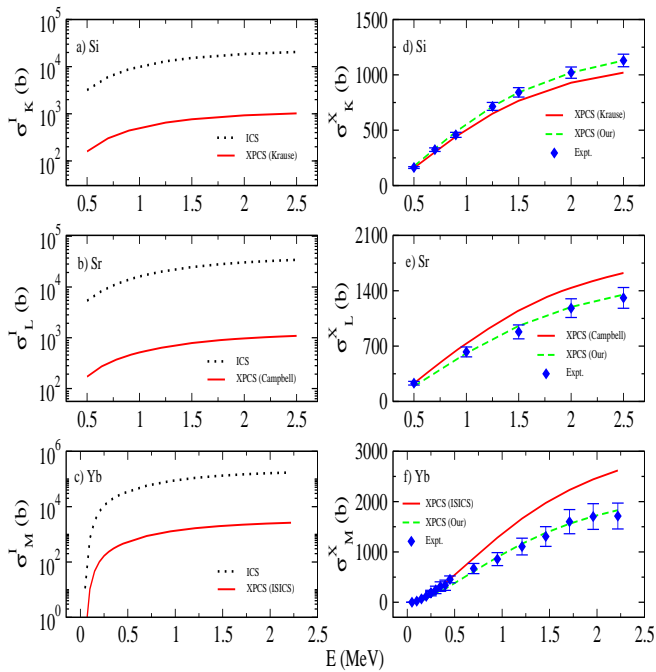}
\vspace{-20pt}
\caption{Nature of ionization cross-sections (ICS) represented with the black dotted line and x-ray production cross-sections (XPCS) with red solid line are quite similar as shown in (a, b, c). Disagreement between experimental and theoretical XPCS with the incorrect fluorescence parameters \cite{campbell2003fluorescence, BATIC20132232} for Si, Sr and Yb shown with a red solid line in Fig. \ref{Match-Mismatch}(d,e,f) have been rectified by using the correct fluorescence parameters as measured by the present method and shown in green dashed line (d, e, f). Note that theoretical XPCS obtained by the fluorescence yields given in \cite{campbell2003fluorescence, BATIC20132232} are shown in (a,b,c) in log scale and that in (d,e,f) are in linear scale.}
\label{modified}
\vspace{-10pt}
\end{figure} 
Atomic parameters include fluorescence yield, CK yield, and fraction of x-ray emission rates. It is obvious from Eqn.\ref{eqn:Sig-Kxx} and Eqn.\ref{eq:L1} that  $\sigma^I_{K}$  and $\sigma_{L_1}$ are proportional to $\sigma^x_{K}$ and $\sigma_{L_{\gamma_{2+3}}}$, respectively. Note that ${\omega_1}{F_{\gamma_{2+3,1}}}$ is an effective fluorescence yield of the $\gamma_{2+3}$ lines. Moreover, the uncertainty in the fraction of emission rates is only 0.2\% \cite{campbell1989interpolated}, so that no appreciable error can be propagated through it. Hence, proportionality constants are nothing but the corresponding fluorescence yield only. Let us consider Eqn.\ref{eq:L2}, which holds a second term on the right-hand side that contains $\sigma_{L_1}f_{12}$. $\sigma_{L_1}$ is of the order of $\sigma_{L_2}$ but $f_{12}$ is a small fraction \cite{singh2022evaluation}, so the second term is much smaller than the first term. In case of Eqn.\ref{eq:L3}, $\sigma_{L_3}$ is much larger than $\sigma_{L_1}$ and $\sigma_{L_2}$; while $f_{ij}$'s are small fractions \cite{singh2022evaluation}. Thus, in all the said cases, ICS is \textcolor{red}{effectively} proportional to XPCS, and the constant of proportionality is nothing but one over the fluorescence yield. This picture can also be shown for M-shell cases. 
\par
\textcolor{red}{Next let us discuss an important point.} Even if we apply total XPCS, $\sigma^x_T=\sum_p\sigma_{L_p}$ and $\sigma^I_T=\sum_i\sigma_{L_i}$, where $\sigma_{L_p}$ and $\sigma_{L_i}$ denote XPCS and ICS for L-shell, respectively, then also $\sigma^x_T$ and $\sigma^I_T$ as a function of energy remain similar to that reflected in Fig.\ref{modified}(a-c) for K-, L- and M-shells, respectively. Therefore, if the fluorescence yield data are right, then a good agreement is observed in Fig.\ref{Match-Mismatch}(a,b,c), otherwise, a departure prevails (Fig.\ref{Match-Mismatch}(d,e,f)) for all K-, L-, and M-shell cases.\\
\begin{figure}
\centering
\includegraphics[width=11cm,height=14.8cm]{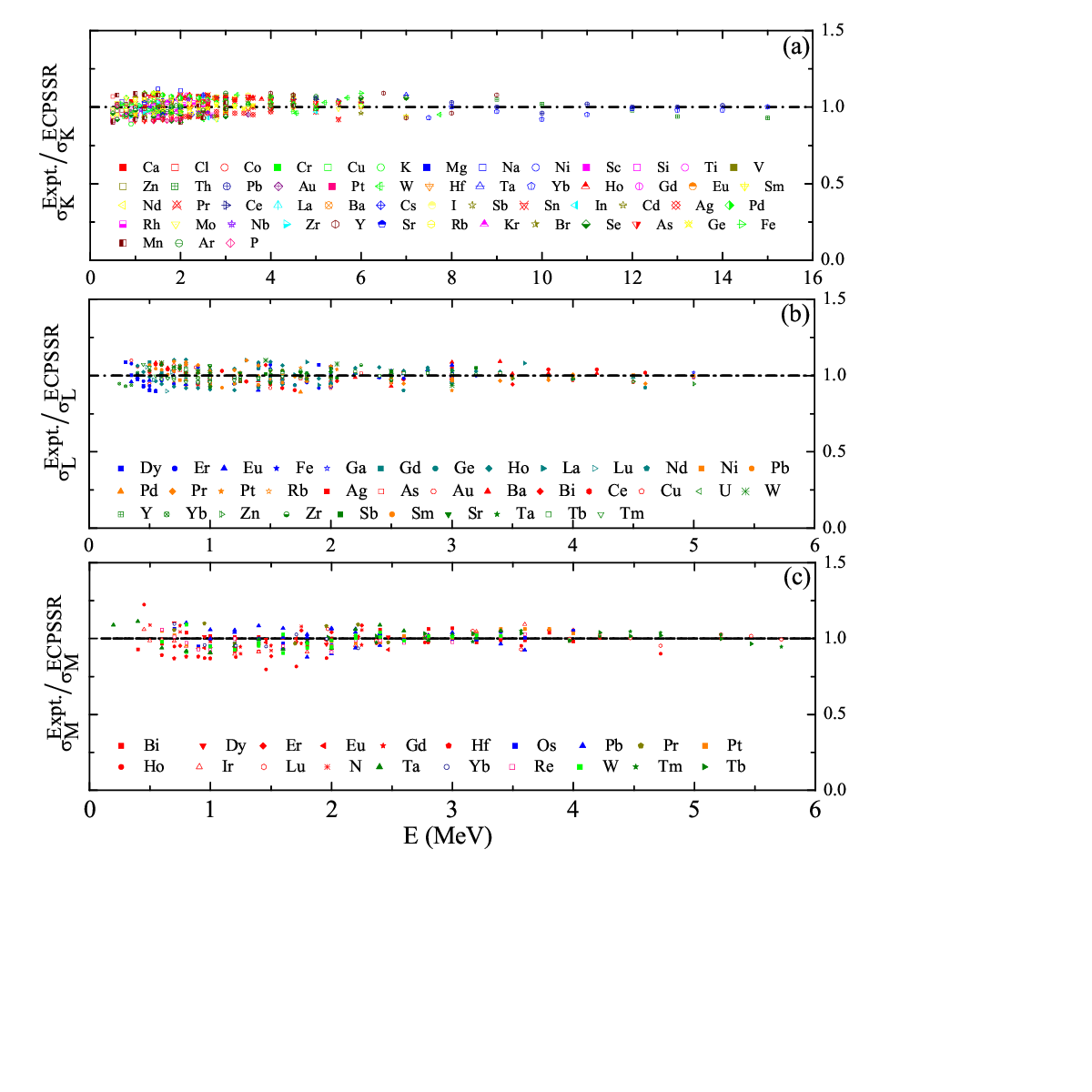}
\vspace{-110pt}
\caption{Agreement achieved between theory and experiment for various target elements is shown using the ratio of experimental to theoretical X-ray production cross-sections versus the proton beam energy for all K-, L- and M-shell cases in (a), (b) and (c), respectively.}
\label{ratio}
\vspace{-10pt}
\end{figure}
\begin{table}[htbt]
\centering
\caption{Comparison of silicon fluorescence yield}
\begin{tabular}{|c|c|c|}
\hline
\textbf{Method} & \textbf{Source} & \textbf{Value} \\
\hline
Experimental & Brunner \cite{brunner1987k} (Mn K$\alpha$)  & 0.0476 $\pm$ 0.0006 \\
  & Brunner \cite{brunner1987k} (Mn K$\beta$)  & 0.0511 $\pm$ 0.0037 \\
 & Campbell \cite{campbell1998experimental}   & 0.051 $\pm$ 0.002 \\
 & Campbell \cite{hopman2012accurate} & 0.0495 $\pm$ 0.0015 \\
 & Present & 0.056 $\pm$ 0.0016 \\
\hline
Theoretical  & McGuire \cite{mcguire1974ionization} & 0.0592 \\
& Walters and Bhalla \cite{walters1971nonrelativistic} & 0.0514 \\
\hline
\end{tabular}
\label{comparison}
\end{table}
\begin{table*}[htbp]
\centering
\caption{Measured fluorescence yield data of some elements are compared with the reported values.} 
\resizebox{16.5 cm}{!}{
\begin{tabular}{|c|c|c|c|c|c|c|c|c|}
\hline
\multirow{2}{*}{Target} & \multicolumn{4}{c|}{K-Shell (Krause \cite{krause1979atomic})} & \multicolumn{4}{c|}{K-Shell (Our)} \\ \cline{2-9} 
                        & Si   & Ti    & Cu    & Ge    & Si    & Ti    & Cu    & Ge    \\ \hline
$w_k$                      & 0.05 & 0.214 & 0.44  & 0.535 & 0.056$\pm$0.002 & 0.214$\pm$0.006 & 0.42$\pm$0.003  & 0.535$\pm$0.0004 \\ \hline

\multirow{2}{*}{Target} & \multicolumn{4}{c|}{L-Shell (Campbell \cite{campbell2009fluorescence})} & \multicolumn{4}{c|}{L-Shell (Our)} \\ \cline{2-9} 
                        & Ga     & Sr     & Ce     & Gd     & Ga     & Sr     & Ce     & Gd     \\ \hline
$w_{1}$                     & 0.0021 & 0.0051 & 0.058  & 0.09   & 0.002$\pm$0.00002 & 0.004$\pm$0.0002 & 0.058$\pm$0.0005 & 0.079$\pm$0.002  \\ \hline
$w_{2}$                     & 0.012  & 0.024  & 0.11   & 0.175  & 0.012$\pm$0.0001  & 0.019$\pm$0.0007  & 0.110$\pm$0.001   & 0.158$\pm$0.004  \\ \hline
$w_{3}$                    & 0.013  & 0.026  & 0.111  & 0.167  & 0.013$\pm$0.0001  & 0.021$\pm$0.0008  & 0.111$\pm$0.001  & 0.155$\pm$0.004  \\ \hline

\multirow{2}{*}{Target} & \multicolumn{4}{c|}{M-Shell (ISICS \cite{BATIC20132232})} & \multicolumn{4}{c|}{M-Shell (Our)} \\ \cline{2-9} 
                        & Yb      & Ir      & Pb      & Bi      & Yb      & Ir      & Pb      & Bi      \\ \hline
$w_{1}$                     & 0.00115 & 0.0018  & 0.00263 & 0.00289 & 0.00075$\pm$0.000008 & 0.0018$\pm$0.00003  & 0.00223$\pm$0.00005 & 0.00289$\pm$0.00006 \\ \hline
$w_{2}$                     & 0.00197 & 0.00353 & 0.00574 & 0.00652 & 0.00128$\pm$0.000014  & 0.00353$\pm$0.00005 & 0.00488$\pm$0.00010 & 0.00652$\pm$0.00013 \\ \hline
$w_{3}$                     & 0.00166 & 0.00356 & 0.00505 & 0.00533 & 0.00108$\pm$0.00001 & 0.00356$\pm$0.00005 & 0.00429$\pm$0.00009 & 0.00533$\pm$0.00011 \\ \hline
$w_{4}$                     & 0.0086  & 0.0171  & 0.03266 & 0.033   & 0.00561$\pm$0.00006 & 0.01710$\pm$0.00025  & 0.02776$\pm$0.00059 & 0.0330$\pm$0.00065   \\ \hline
$w_{5}$                     & 0.0149  & 0.02377 & 0.03094 & 0.0325  & 0.0097$\pm$0.0001 & 0.02377$\pm$0.00035 & 0.0263$\pm$0.00056 & 0.0325$\pm$0.00064  \\ \hline
\end{tabular}}
\label{fluorecence}
\end{table*}
If the ECPSSR calculation does not agree with the experiment, \textcolor{blue}{we vary the fluorescence yield until a good consonance is found, and this way we accurately determine the fluorescence yields. Our initial guess for this iteration is always the reported values used in the ISICSoo code. We vary all the fluorescence yield data equally; it means that we change each one with the same percentage} To employ this method, we accept only the set of experimental proton-induced XPCS data that is available for a wide energy range. Such data can be taken from a single experiment or from multiple experiments if a smooth transition between two data sets persists. This procedure constitutes an experimental method \textcolor{red}{powerful} for measuring the fluorescence yields. The fluorescence yield data obtained from this method transform the disagreements [Fig.\ref{Match-Mismatch}(d-f)] into an example of good agreement with the theory [Fig. \ref{modified} (d-f)]. In this way, we have found the accurate fluorescence yield data sets for the elements where proton-induced experimental XPCS data are available. The current scenario of closeness between theory and experiment is shown by means of a ratio of experimental to theoretical XPCS versus proton beam energies for many target elements. The large scattering seen with the ratios for K- and L-shell \cite{lapicki2005status,lapicki2020status} has disappeared as shown in Fig.\ref{ratio}. This figure shows a very good agreement between theory and experiment even for M-shell too. Moreover, this achievement opens the way to making proton-induced PIXE more reliable.
\par
\textcolor{red}{We get from Eqn.\ref{eqn:Sig-Kxx} that $\omega_k$, $\frac{\Delta\omega_k}{\omega}$ is equal to $\frac{\Delta\sigma^x_k}{\sigma^x_k}$ if we assume $\sigma^I_k$ obtained from the ECPSSR theory does not introduce considerable uncertainty compared to the uncertainty of the measurement in $\sigma^x_k$. This means that the measurement uncertainty of the K-shell fluorescence yield can be evaluated from the measurement uncertainty of $\sigma^x_k$. Recall that the fluorescence yields are not energy-dependent, but the XPCS is. To get energy independent measurement uncertainty for the $\sigma^x_k$, we define $\frac{\Delta\omega_k}{\omega}$ as follows}
\begin{equation}
\frac{\Delta\omega_k}{\omega_k} =\frac{\sqrt{\sum_i^n [y_i - f(E_i)]^2}}{\sum_i^n {y_i}}.
\label{uncertainty}
\end{equation}
where $y_i$ is experimental $\sigma^x_{k}$ at energy $E_i$, $f(E_i)$ is theoretical $\sigma^x_{k}$ at energy $E_i$ and n is the number of data points available with the experiment.  This method of finding the uncertainty is inapplicable for the L-shell because three fluorescence yields act in this case. \textcolor{red}{Furthermore, as already mentioned we take an initial guess of fluorescence yields used in the ISICSoo code and we vary each one with the same percentage. It implies the following relation for the L-shell case}
\begin{gather}
\frac{\Delta\omega_1}{\omega_1} = \frac{\Delta\omega_2}{\omega_2} = \frac{\Delta\omega_3}{\omega_3} = \frac{\Delta\omega}{\omega}.
\end{gather}
Details of the error calculations are given in Appendix A. Obviously, a similar equation will also apply to the M-shell too. 

\begin{figure}
\centering
\includegraphics[width=8.7cm,height=9cm]{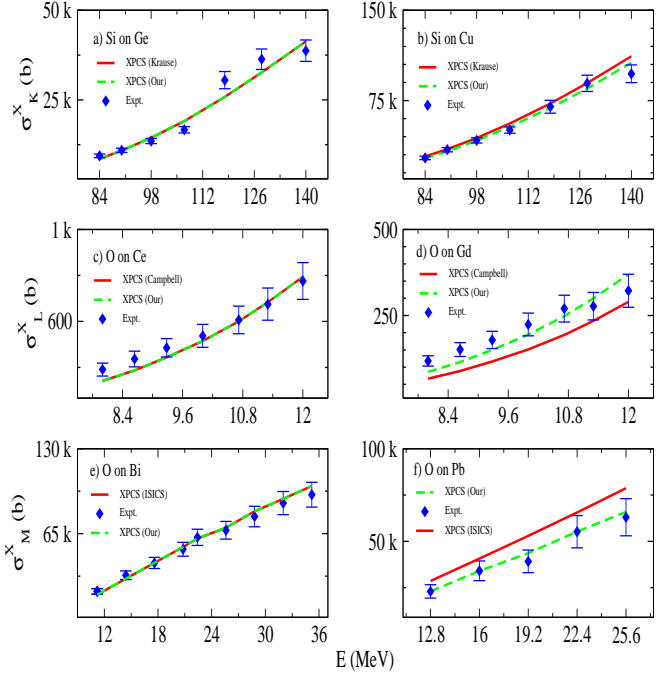}
\vspace{-20pt}
\caption{Heavy ion induced K, L, and M X-ray production cross-sections are displayed as a function of energy in (a,b), (c,d), and (e,f), respectively to compare theory (green dashed line) using the present fluorescence yield data with experiment. Red lines show the comparison between experiment and theory with the fluorescence yield data of \cite{campbell2003fluorescence, BATIC20132232}.}
\label{heavyion}
\vspace{-10pt}
\end{figure} 

The present method for measuring the fluorescence yields is based on proton-induced XPCS measurements as a function of beam energy, and thus it is not restricted to any element. Hence, it can be treated as a universal technique. In the past, different methods have been used to measure the fluorescence yields, and we present a comparative study of silicon fluorescence yield investigations. The comparison given in Table \ref{comparison} shows a wide range of values. We find a good agreement between our measurement and that of \citet{brunner1987k} using the escape peak of Mn $K_\beta$ and in the middle of two theoretical works \cite{walters1971nonrelativistic,mcguire1974ionization}. Other experimental works report smaller values. Importantly, our method uses several energy points to find an average value of the fluorescence yield, which is reliable. As a representative, we show a comparison of the fluorescence yields between the present values with the previously reported ones \cite{krause1979atomic, campbell2003fluorescence, BATIC20132232} in Table \ref{fluorecence} for the elemental targets utilized here. All fluorescence yield data will be reported elsewhere.\\

\indent
With the fluorescence data determined, we proceed to verify the current scenarios based on our understanding of the heavy-ion-induced ionization data. Here, we apply the theoretical method used for the M-shell in our recent paper \cite{kaur2023understanding} to take multi-ionization and vacancy creations in the target atoms due to electron capture by the heavy projectile ions not only for M-shell but it has also been adopted for the K- and L-shell too. To test this method, we take two cases each for K-, L-, and M-shells for comparison with theoretical calculations, as shown in Fig.\ref{heavyion}. Interestingly, here we find also that a remarkable agreement is achieved throughout. This development not only solves a fundamental physics issue but also overcomes all the barriers to using the HIPIXE method. \\
\section{Conclusion}
\indent To conclude, the basic issue with the understanding of heavy-ion-induced collision physics is the incorrect fluorescence yields, which severely affect the estimation of the DCI including multi-ionization effects.  Still, it has never received significant attention. Another important issue is the electron capture-induced target ionization, which has been solved recently \cite{chatterjee2021significance,chatterjee2022understanding,kaur2023understanding}. \textcolor{red}{Interestingly, it has been possible now to find reliable fluorescence yields within certain measurement uncertainties that come from the corresponding experimental XPCS for a wide range of energy. This has been possible because of the availability of accurate proton-induced ionization cross sections using the ECPSSR theory (accurate in the sense that the theoretical values agree with the experimental ones within the measurement uncertainties)} and the theoretical method reported recently to estimate the electron capture-induced vacancy production in target inner shells \cite{kaur2023understanding} enable us to move towards a full understanding of heavy-ion-induced collision physics. Currently, this success gives the way to making both LIPIXE and HIPIXE more reliable and accurate. 
\section{Acknowledgements}
It is our great pleasure to acknowledge C.C. Montanari for providing the proton-induced ionization cross sections using the Shellwise Local Plasma Approximations.   

\appendix
\section{Details of the error calculations}

Theoretically K x-ray production cross section (XPCS) $\sigma^x_{K}$ is related to K-shell ionization cross section ($\sigma_K^I$) as follows
\begin{equation}
\sigma^x_{K}= {\omega_K}\sigma^x_{K}\label{eqn:Sig-Kx}
\end{equation}
where $\omega_K$ is the K x-ray fluorescence yield. 
\par
Theoretical $L$ X-ray production cross sections for the most commonly resolved $L_l$, $L_{\alpha}$, $L_{\beta}$, $L_{\gamma}$ x-rays are related to the $L_i$ subshell ionization cross sections, $\sigma_{Li}$, as follows
\begin{gather}
\sigma^{x}_{L_l}=[\sigma_{L_1}(f_{13}+(f_{12}f_{23})+\sigma_{L_2}f_{23}+\sigma_{L_3}]\omega_3F_{3l} \,,
\label{eq:1}
\end{gather}
\begin{gather}
\sigma^{x}_{L_{\alpha}}=[\sigma_{L_1}(f_{13}+(f_{12}f_{23})+\sigma_{L_2}f_{23}+\sigma_{L_3}]\omega_3F_{3\alpha} \,,
\label{eq:2}
\end{gather}
\begin{gather}
\begin{split}
\sigma^{x}_{L_{\beta}}=\sigma_{L_1}[\omega_1F_{1\beta}+f_{12}\omega_2F_{2\beta}+(f_{13}+(f_{12}f_{23})\omega_3F_{3\beta}]+\\
\sigma_{L_2}(\omega_2F_{2\beta}+f_{23}\omega_3F_{3\beta})+\sigma_{L_3}\omega_3F_{3\beta} \,,
\label{eq:3}
\end{split}
\end{gather}
and 
\begin{gather}
\sigma^x_{L_{\gamma}}=\sigma_{L_1}\omega_1F_{1\gamma}+(\sigma_{L_1}f_{12}+\sigma_{L_2})\omega_2F_{2\gamma} \,
\label{eq:4}
\end{gather}
Here $\sigma_{L_p}^x$s are the x-ray production cross sections of the different $L$ x-ray components (${L_p}$, $p =l,\,\alpha,\,\beta,\,\gamma$).  $\sigma_{L_i}$ is the ionization cross sections for the $L_i$ subshell with $i= 1,\,2,\,3$. $\omega_i$ is the fluorescence yield of the $L_i$ subshell, $f_{ij} (i < j)$ is the CK yield for the CK transition between the $L_i$ and $L_j$ subshells with $j= 1,\,2,\,3$, and $F_{ip}$ is the fractional radiative emission rates.
\begin{gather}
\begin{split}
\sigma^x_{T}= \sum_P\sigma^x_{L_P} = \omega_1 [\sigma_{L_1}F_{1\gamma} + \sigma_{L_1}F_{1\beta}] +\\ 
\omega_2 [(\sigma_{L_1}f_{12}+\sigma_{L_2})F_{2\gamma} + \sigma_{L_2}F_{2\beta} + \sigma_{L_1}F_{1\beta}] +\\ 
\omega_3 [\sigma_{L_3}F_{3\beta} + f_{23}F_{3\beta}\sigma_{L_2} + \sigma_{L_1}F_{3\beta}(f_{13}+(f_{12}f_{23}) +\\
[\sigma_{L_1}(f_{13}+(f_{12}f_{23})+\sigma_{L_2}f_{23}+\sigma_{L_3}]F_{3l}]
\end{split}
\end{gather}
Let us see now how the uncertainty in $\sigma^x_{T}$ is related to that of $\omega_i$s.
\begin{gather}
\begin{split}
\frac{\Delta\sigma^x_T}{\sigma^x_{T}} = \frac{\Delta\omega_1}{\omega_1}\times\frac{\omega_1}{\sigma^x_{T}}[\sigma_{L_1}F_{1\gamma} + \sigma_{L_1}F_{1\beta}] +\\ 
\frac{\Delta\omega_2}{\omega_2}\times\frac{\omega_2}{\sigma^x_{T}} [(\sigma_{L_1}f_{12}+\sigma_{L_2})F_{2\gamma} + \sigma_{L_2}F_{2\beta} + \sigma_{L_1}F_{1\beta}] +\\ 
\frac{\Delta\omega_3}{\omega_3}\times\frac{\omega_3}{\sigma^x_{T}}[\sigma_{L_3}F_{3\beta} + f_{23}F_{3\beta}\sigma_{L_2} + \sigma_{L_1}F_{3\beta}(f_{13}+(f_{12}f_{23}) +\\
[\sigma_{L_1}(f_{13}+(f_{12}f_{23})+\sigma_{L_2}f_{23}+\sigma_{L_3}]F_{3l}]
\end{split}
\end{gather}
Suppose 
\begin{gather}
\frac{\Delta\omega_1}{\omega_1} = \frac{\Delta\omega_2}{\omega_2} = \frac{\Delta\omega_3}{\omega_3} = \frac{\Delta\omega}{\omega}
\end{gather}
Then we find that
\begin{gather}
\frac{\Delta\sigma^x_T}{\sigma^x_{T}} = \frac{\Delta\omega}{\omega}
\end{gather}
We know that the $\sigma^x_{T}$ is beam energy dependent but $\omega_i$s are not. To find out the energy independent uncertainty in $\omega_i$s, we define it as follows
\begin{equation}
\frac{\Delta\sigma^x_{T}}{\sigma^x_{T}} = \frac{\Delta\omega}{\omega} =\frac{\sqrt{\sum_i^n [y_i - f(E_i)]^2}}{\sum_i^n {y_i}},
\end{equation}
where $y_i$ and $f(E_i)$ are experimental and theoretical $\sigma^x_{T}$ at $E_i$ with the measured $\omega_i$s in the present work.  Similarly, we can show that such a relation is valid for the M-shell ionization too.

\bibliography{BIB.bib}
\bibliographystyle{apsrev4-1}
\end{document}